# Short-term heterologous immunity after severe influenza A outbreaks

S. Towers[1*], Z. Feng[1], & N. Hupert[2]

**Conventional wisdom holds that influenza A and B are such genetically dissimilar viruses that infection with one cannot confer cross-immunity to the other. However, our examination of the records of the past 25 influenza seasons in the U.S. reveals that almost every time there is an early and severe influenza A outbreak, the annual influenza B epidemic is almost entirely suppressed (and is never suppressed otherwise). Temporary broad-spectrum (aka "heterologous") immunity in the aftermath of influenza infection is the most direct explanation for this phenomenon. We find a remarkably weak degree of temporary cross-immunity is needed to explain these patterns, and that indeed influenza B provides an ideal setting for the observation of heterologous immune effects outside of the carefully controlled environment of a laboratory.**

Our examination of the past 25 U.S. influenza seasons reveals that annual influenza B epidemics are suppressed only when they are preceded by an early and severe influenza A season. These observed patterns suggest that at least some level of cross-immunity to influenza B is conferred by prior infection with influenza A, despite conventional wisdom that this is not the case due to the many genetic differences between the two viruses (see, for instance, [1] and [2]). In recent years compelling evidence of temporary broad-spectrum (also known as "heterologous") cross-immunity among completely unrelated viruses has been observed in carefully controlled animal experiments in a laboratory setting [3]-[5]. However, up until now observation of these effects outside of a clinical setting has proved elusive, indicating that the level of cross-immunity conferred is likely weak. Here we calculate the level of broad cross-strain immunity necessary to produce the observed epidemiological patterns in the strain-specific influenza attack rates and we find that a remarkably small degree of temporary heterologous immunity is needed. We also find that the epidemiological properties of influenza provide an almost ideal setting for observation of these weak effects, and that influenza appears to be rather unique among common diseases in this respect. Thus the influenza epidemiological patterns of the past 25 years not only provide an indirect observation of heterologous immune effects outside of the laboratory, but also allow us to quantify of the degree of such immunity conferred by prior viral infection, contributing considerably to the understanding of immunity in the aftermath of infection and recovery from viral disease.

[1] Mathematics Department, Purdue University, West Lafayette, IN 47907

[*] stowers@purdue.edu

[2] Weill Medical College, Cornell University, New York, NY 10065

Influenza viruses come in three varieties--A, B, and C--of which the first causes the majority of morbidity and mortality in humans. Influenza A not only slowly mutates over time (antigenic shift), but also re-assorts with animal influenza A viruses, such as those that circulate among birds and swine; these two mechanisms prevent the development of lasting humoral immunity to influenza A infection, and are the main reason that influenza A causes occasional global pandemics. In the past 25 years two main types of influenza A have circulated in the population, A(H1N1) and A(H3N2), with many antigenically diverse subvariants. Influenza B generally circulates only in humans and seals, and largely due to this it can only mutate via antigenic shift, resulting in less genetic diversity in circulating strains compared to influenza A. Due to this lack of genetic diversity, humans usually acquire a partial lifelong immunity upon their first infection. For this reason influenza B does not cause pandemics, and is also usually less transmissible in the population compared to influenza A. Conventional wisdom holds that influenza A cannot confer immunity to influenza B because they are such genetically different viruses. Influenza C is a generally mild disease, and is of no epidemiological importance.

Ample evidence exists in animal studies for cross-immunity between different subtypes of influenza A, such as A(H1N1) conferring immunity to A(H3N2) and/or vice-versa[6]-[9], and the phenomenon has even been observed in humans[10]. Models of influenza dynamics with cross-immunity between A strains have been developed and studied (see, for instance [11]-[13]). However, this kind of within-type cross-immunity does not explain why influenza B was almost entirely suppressed in the great majority of seasons where a strain of influenza A peaked early, as seen in Figure 1; in the last 25 U.S. influenza seasons between 1985-86 to 2009-10, eleven have had an influenza A epidemic with an early peak occurring before the end of January, and eight of those eleven seasons saw an almost total suppression of the usual B epidemic (i.e., influenza B accounted for ≤ 1% of the total influenza isolates during those 8 seasons).[†] The remaining 14 seasons had an influenza A epidemic that peaked after the end of January, and influenza B was not suppressed during any of those seasons and accounted for at least 12% of the influenza isolates in each of those years. The hypergeometric probability that this observed temporal pattern occurred just by random chance is $p = 0.0002$.

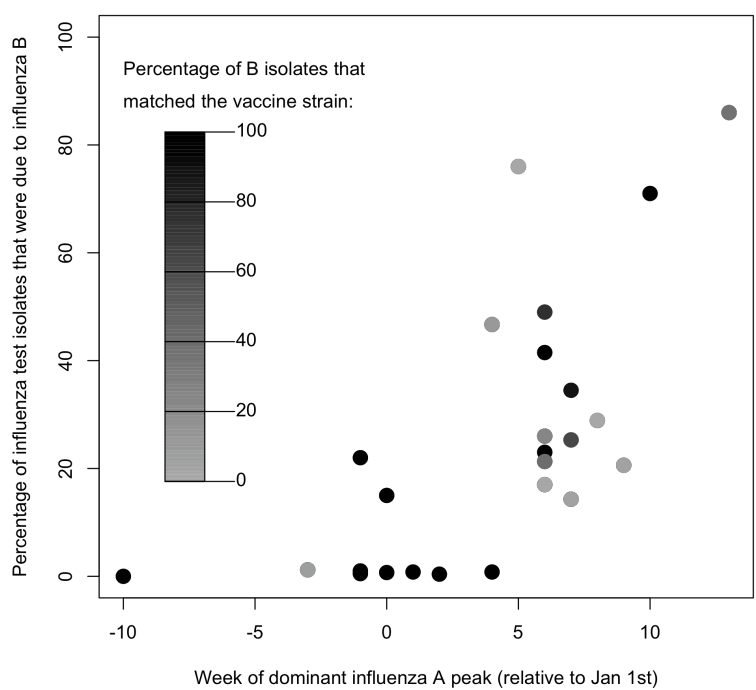

**Figure 1: Summary of test isolate data of the 1985-86 to 2009-10 U.S. influenza seasons.**

---

[†] A citation list of publicly available summaries of the past 25 U.S. influenza seasons is included in the supplementary information. The eight seasons out of the past 25 where influenza B was suppressed after an early influenza A epidemic are 1986-87 A(H1N1), 1989-90 A(H3N2), 1991-92 A(H3N2), 1993-94 A(H3N2), 1997-98 A(H3N2), 1999-00 A(H3N2), 2003-04 A(H3N2) and, notably, the recent 2009-2010 A(H1N1) season. The 1995-95, 1996-97, and 2000-01 seasons had an early A epidemic followed by a seasonal B epidemic.

These temporal patterns do not appear to be due to occasional mismatches between the influenza B vaccine composition and the dominant circulating strain. When we examine the inter-correlations between the timing of the dominant influenza A peak, the fraction of all isolates each season due to influenza B, and the fraction of B isolates each season that matched the B vaccine strain, we find that only the first two are significantly correlated; when we examine the partial correlations between these three, we also find that only the first and second, controlled for the third, are significantly correlated.[‡] We also find that the timing of the dominant influenza A peak is not significantly correlated to the size of the influenza B epidemic in the next season. We thus conclude that the timing of the influenza A peak is the apparent dominant factor in the dynamics of the ensuing B influenza season, but that the effect is only temporary.

Thus, while conventional wisdom and published reports hold that influenza A cannot confer immunity to influenza B, the observations of the past 25 influenza seasons strongly suggest that highly infectious strains of influenza

---

[‡] Let us refer to the timing of the influenza A peak, the fraction of all isolates each season due to influenza B, and the fraction of B isolates each season that matched the B vaccine strain, as X, Y, and Z, respectively. The correlations between X and Y, Y and Z, and X and Z are +0.67 (two-tailed p = 0.0002), -0.29 (p = 0.16), and -0.38 (p = 0.06), respectively. The partial correlations between X and Y given Z, Y and Z given X, and X and Z given Y are +0.63 (two-tailed p = 0.0007), –0.05 (p = 0.8), and –0.27 (p = 0.2), respectively. None of the inter-correlations between X and the next season's Y and Z are significant.

A do indeed confer at least some temporary broad-spectrum heterologous immunity to influenza B. A relatively recent body of work has examined heterologous cross-immunity[3]-[5][14], and has come to the conclusion that short term broad-spectrum immunity conferred by previous viral infections plays an important role in the functioning of the immune system. This kind of cross-immunity has in fact been observed in the laboratory for completely unrelated viruses; for instance, the study presented in reference [3] finds that infection with influenza A appears to stimulate short term immunity to hepatitis C in mice. However, little is known about the degree of heterologous immunity conferred by viral infection.

To calculate the degree of temporary broad-spectrum immunity needed to explain the observed suppression of influenza B epidemics, we examine the scenario where a fraction $q$ of the population has been infected with influenza A, subsequently conferring broad-spectrum immunity to a fraction $\alpha$ of those people. In the immediate aftermath of an influenza A epidemic, a fraction $v$ of the population thus has vaccination immunity to influenza B, and a further fraction $(1-v)q\alpha$ has broad-spectrum immunity conferred by the infection with influenza A. The total "herd immunity" fraction in this case is thus $p = v + (1-v)q\alpha$.

If $p$ is greater than a certain threshold, the disease will not be able to spread within the population (for instance, smallpox was eradicated by vaccinating a large enough fraction of the population to achieve sufficient herd immunity). The less infectious a disease is, the smaller the threshold on $p$.

Influenza is not very infectious compared to most other diseases with significant population-wide morbidity and mortality, and thus the threshold value of $p$ is fairly low, and the cross-immunity $\alpha$ thus does not need to be very large in order to suppress an influenza epidemic in the scenario described above. As described in Materials and Methods in the supplementary information, we use the epidemiological properties of influenza and current vaccination patterns to calculate the limit on $\alpha$ needed to suppress influenza B after an early and severe influenza A epidemic, and find that $\alpha$ above 0.01 to 0.31 is sufficient. This is considered very weak to weak cross-immunity [13].

We thus conclude that the observed influenza epidemiological patterns of the past 25 years not only provide evidence of heterologous immunity effects occurring outside of the carefully controlled environment of a laboratory*, but also, rather remarkably, allow us to quantify the degree of such immunity conferred after viral infection. Our results suggesting temporary low-level heterologous immunity between influenza A and B are also reflected in the shape of the human influenza phylogenic tree; references [15]-[17] point out that large scale modeling cannot replicate the influenza phylogenic tree structure unless broad-spectrum temporary immunity (lasting one or two months) after influenza infection is included in the model. In fact, the analysis in [15] estimates that the magnitude of this broad immunity could be very weak (within the range of α = 5% to 20%) and still reproduce the observed influenza phylogenic structure. This is in excellent agreement with the results we derive here. There may in fact be a grain of truth in the old adage "that which does not kill you makes you stronger", if indeed only slightly (and temporarily).

---

* It should be pointed out here that while heterologous immune effects have been observed in the laboratory for other diseases, there is a surprising dearth of animal or human influenza challenge studies where infection with influenza A is shortly followed by challenge with influenza B.

**Acknowledgements** Z.F. gratefully acknowledges the partial support of this work by National Science Foundation Grant DMS-0719697. N.H. is grateful for the support of an unrestricted grant from the Stafford Family Foundation to the Weill Cornell Institute for Disease and Disaster Preparedness.

## Supplementary information

### Materials and Methods

Summaries of past 25 influenza seasons in the U.S. were obtained from references [31]-[43]. Additional information about year-to-year influenza vaccine composition was obtained from references [44] and [45].

To calculate the degree of temporary broad-spectrum immunity needed to explain the observed suppression of influenza B epidemics after early and severe influenza A epidemics, we examine the scenario where a fraction $q$ of the population has been infected with influenza A, subsequently conferring broad-spectrum immunity to a fraction $\alpha$ of those people. We also assume that a fraction $v$ of the population already has vaccination immunity to influenza B, and that $q$ and $v$ are independent (i.e., that prior vaccination for influenza B neither increases nor decreases one's chances for catching influenza A). In the immediate aftermath of an influenza A epidemic, a fraction $v$ of the population thus has vaccination immunity to influenza B, and a further fraction $(1-v)q\alpha$ has broad-spectrum immunity conferred by the infection with influenza A. The total "herd immunity" fraction in this case is thus $p = v + (1-v)q\alpha$.

The reproduction number, $R_0$, of a disease is the mean number of secondary cases a typical single infected case will generate in a population with no immunity to the disease. If a fraction $p$ of the population has prior immunity to the disease, then the $R_0$ of the disease is effectively reduced to $R_{eff} = R_0(1-p)$.

As soon as $R_{eff} < 1$, the disease will die out in the population. In order to fully suppress the B epidemic, we thus must have $(1-p) < 1/R_0$, which implies

$$q\alpha > 1 - \frac{1}{R_0(1-v)} \quad . \quad (1)$$

Notice that $R_0(1-v)$ is the effective $R_0$ due to vaccination alone. Let us refer to this as $R_{vacc}$, and re-write Equation 1 as

$$\alpha > \frac{R_{vacc} - 1}{qR_{vacc}}. \quad (2)$$

It is now obvious that the closer $R_{vacc}$ is to 1, the smaller $\alpha$ needs to be to completely suppress an influenza B epidemic. In fact, if the fraction of people, $q$, infected by influenza A is large (as it is in a pandemic situation), the size of $\alpha$ needed to suppress B can be quite small indeed.

The average $R_0$ for seasonal influenza A and B has been estimated to be between roughly 1.1 to 1.7, with an average around 1.3 to 1.4 [46]-[48]. Influenza B tends to be less infectious than influenza A, thus it is thought to have an $R_0$ at the lower end of this range[48]. In this analysis we will assume the $R_0$ of influenza B is 1.2 to 1.3. Typical influenza vaccination coverage in the US is around 23%, but it is estimated that fewer than 50% to 70% of vaccinated people actually achieve full immunity[49]-[52]. We thus assume a range of $v$ between 11% to 16%[48]. Since $R_0 = 1.2$ to 1.3, and $v = 11\%$ to 16%, we find that the effective $R_0$ due to vaccination alone is $R_{vacc} = 1.01$ to 1.16. To estimate $q$, we note that the $R_0$ of A(H1N1) has been measured to be between 1.4 and 1.6 (i.e., larger than that of seasonal influenza)[53], and modeling studies using different methodologies have estimated that roughly 45% to 70% of the population were infected with A(H1N1) by the end of 2009[54][55]. This is in agreement with CDC estimates that 14% to 28% of the population fell ill[56], when it is taken into account that only approximately 40% of those infected are symptomatic[57]. Further, the CDC estimates that around 20% of the population falls ill during a severe influenza season, thus we see that the 2009-10 season is roughly comparable to the other influenza seasons where a highly infectious strain of influenza A caused an early peak.

Substituting these numbers for $R_0$, $v$, and $q$ into Equation 2 yields that $\alpha$ must be greater than around 0.01 to 0.31 to fully suppress a B epidemic in a population with typical vaccination levels. This level of cross-immunity is considered to be very weak to weak[16]. Using the cross-immunity disease model in [16], we can simulate the dynamics of competition between influenza strains with this low level of cross-immunity, and find that it does not significantly impact the progression of either epidemic when they occur simultaneously. This conclusion is supported by the studies presented in [58].